Direct-Current Induced Dynamics in $Co_{90}Fe_{10}/Ni_{80}Fe_{20}$ Point Contacts

W. H. Rippard, M. R. Pufall, S. Kaka, S. E. Russek, and T. J. Silva

*National Institute of Standards and Technology,* Boulder, CO 80305

Abstract:

We have directly measured coherent high-frequency magnetization dynamics in ferromagnet films induced by a spin-polarized DC current. The precession frequency can be tuned over a range of several gigahertz, by varying the applied current. The frequencies of excitation also vary with applied field, resulting in a microwave oscillator that can be tuned from below 5 GHz to above 40 GHz. This novel method of inducing high-frequency dynamics yields oscillations having quality factors from 200 to 800. We compare our results with those from single-domain simulations of current-induced dynamics.

 PACs Codes: 75.47.-m, 75.75.+a, 85.75.-d

Since the initial predictions of Slonczewski [1] and Berger [2] that a spin-polarized current can induce magnetic switching and dynamic excitations in ferromagnetic thin films, a great deal of work has focused on understanding the interactions between polarized currents and ferromagnetic nanostructures.[3] It was predicted, and later confirmed, that this effect can lead to current-controlled hysteretic switching in magnetic nanostructures in moderate applied magnetic fields.[4,5] This behavior is not only of scientific interest but also finds potential applications in devices such as current controlled switching of magnetic random access memory elements and has implications for the stability of magnetic hard-disk read heads. Another prediction is that the spin-torque can drive steady-state magnetization precession in the case of applied fields large enough to oppose hysteretic switching.[1,2] Numerous device applications exist for such current-controlled microwave oscillators that are integrable with semiconductor electronics.[6] However, with one recent exception in nanopillar devices,[7] to date no direct measurements of these high-frequency dynamics have been reported.[4,5,8] Here we report direct measurements of spin-torque induced magnetization dynamics for both in-plane and out-of-plane applied fields as a function of field strength $H$ and current $I$, and compare the results with simulations based on the theoretical model presented in Ref. [1].

The studies discussed here were carried out on lithographically defined point contacts to spin-valve mesas (8 μm x 12 μm). The point contacts are nominally 40 nm diameter circles, have resistances between 4 Ω and 10 Ω, and show no indications of tunneling in their current-voltage characteristics. Top and bottom electrical contacts to the devices are patterned into a 50 Ω coplanar waveguide, to minimize reflection of high-frequency signals as they propagate off-chip. Fabrication details will be discussed elsewhere.

The spin-valve structures are a (2.5 nm)/Cu (50 nm)/Co$_{90}$Fe$_{10}$ (20 nm)/Cu (5nm)/Ni$_{80}$Fe$_{20}$ (5 nm)/Cu (1.5 nm)/Au (2.5 nm) and show typical giant-magnetoresistance (GMR) values of 80 mΩ. The Co$_{90}$Fe$_{10}$ is considered the "fixed" layer in terms of the spin-torque effect due to its larger volume, exchange stiffness, and saturation magnetization compared with Ni$_{80}$Fe$_{20}$.[9] The device is contacted with microwave probes and a DC current is injected through a bias-tee, along with a 20 µA AC current (500 Hz), allowing simultaneous measurement of the DC resistance, differential resistance, and microwave response. The devices are current-biased so that changes in the relative angle between the Ni$_{80}$Fe$_{20}$ and Co$_{90}$Fe$_{10}$ layers appear as voltage changes across the point contact. The high-frequency voltage signal is amplified and measured with either a 50 GHz spectrum analyzer or a 1.5 GHz real-time oscilloscope. The bandwidth of the off-chip circuitry is 0.1 GHz to 40 GHz. Measurements were performed at room temperature. All results discussed below occur for only one direction of current, corresponding to electrons flowing from the top contact into the spin-valve.

Figure 1a shows a differential resistance d$V$/d$I$ curve of a point contact taken with an in-plane applied field $\mu_0 H = 0.1$ T. The non-hysteretic peak in the d$V$/d$I$ curve, at $I$ = 4 mA in Fig. 1a, has been taken as indirect evidence of magnetization dynamics induced by the spin-torque.[4,5,8] Tsoi *et al*. demonstrated changes in the DC transport properties of point contacts under the influence of external radiation, implying a relationship between spin dynamics and DC resistance.[8] Here we observe these dynamics directly, as shown by the spectra in Fig. 1b. For low currents, no peaks are observed in the frequency spectra. As $I$ is increased to 4 mA ($\approx$ 3 x $10^8$ A/cm$^2$) a peak is observed at $f$ = 7.9 GHz. As $I$ is further increased, the frequency decreases, a trend observed for all in-plane fields measured. This frequency "red shift" is linear in current (see inset) and typically varies from $\approx$ 0.2 GHz/mA at low fields ($\approx$ 50 mT) to $\approx$ 1 GHz/mA at fields of $\approx$ 0.8 T. At higher values of $I$, the excitations decrease in magnitude until no high frequency peaks are observed, as shown in the $I$ = 9 mA spectrum. Assuming the high-frequency signal results from a GMR response, we estimate a maximum excursion angle between the layers of approximately 20 degrees. Because the measured peak amplitude does not increase linearly with current (as it would for fixed excursion angle), we infer that the orbit traversed by the magnetization changes with $I$. The dynamics shown here are strongly correlated with the peak in the d$V$/d$I$ curve. This is not the case for all devices: Typically the onset of the dynamics occurs only in the vicinity of a feature (step, peak, or kink) in d$V$/d$I$, and the relative position of this onset varies with field.

To better understand the possible trajectories of these excitations, we compare our results with simulations that assume an isolated single-domain particle (40 nm x 40 nm) whose behavior is described by a modified Landau-Lifshitz-Gilbert equation proposed by Slonczewski.[1] Finite-temperature effects are included through a randomly fluctuating applied field.[10] The model only approximates the point contact geometry, in which the region undergoing current-induced excitations is coupled to a continuous film by intralayer exchange. For example, effects associated with the formation of domain walls between the region under the contact area and the rest of the film are not included, nor are the effects of spin-wave radiation damping.[1]

The simulations show two basic regimes of motion for in-plane fields. At low current, when oscillations begin, the magnetization *M* precesses in a nearly elliptical mode about the applied field. Here, the time-averaged magnetization <*M*> is parallel to *H*. As *I* increases, the trajectories become non-elliptical and have greater excursion angles with respect to *H*. However, *M* continues to precess around the applied field, while <*M*> changes from parallel to antiparallel alignment with *H*. Within this regime, the simulated excitation frequency decreases approximately linearly with *I*, in agreement with the data shown in Fig. 1b. In this regime, d*f*/d*I* increases with increasing *H*, also in agreement with our measurements. As *I* is further increased the second regime is reached and the simulations show *M* precessing out-of-plane with the precession frequency increasing with current. Consequently, we infer that the observed excitations correspond only to in-plane precession. The out-of-plane precessional regime may not be accessed in our experiments due to a lack of stability of the trajectories in our devices or because the devices are unable to support sufficient current densities. It may also indicate a need to incorporate micromagnetic effects in the modeling.

The measured peak widths are quite narrow. The peaks in Fig. 1b have full-widths-at-half-maximum (FWHM) of ≈ 20 MHz and quality factors $Q = f/$(FWHM) of ≈ 350, with particular values depending on *I*. The FWHMs of the excitations are a weak function of field, leading to values of $Q > 500$ for $f > 30$ GHz. These narrow linewidths indicate that the excitations can be considered coherent single-mode oscillations. Analogous linewidths in ferromagnetic resonance (FMR) measurements would give damping parameters of $\alpha = 1 - 5 \times 10^{-4}$, with the particular value depending on *H*.[11] Our modeling requires an $\alpha = 0.5 - 1 \times 10^{-3}$ to produce similar linewidths at 300 K.

Either analysis gives values of $\alpha$ much smaller than values obtained through field-induced excitations of $Ni_{80}Fe_{20}$ thin films ($\alpha = 0.01$ to $0.005$).[12,13] Linewidths we have measured in nanopillar devices (not shown here) are about a factor of five larger than those measured in point contacts, indicating that the narrowness of these peaks is not a general result for current-induced excitations. The lack of physical magnetic edges in point contact devices may account for their narrow linewidths in comparison to nanopillars. Increased linewidths and effective damping are often found in magnetic nanostructures, resulting from $M$ at the edges of patterned devices lagging $M$ at the center of the device during large-angle oscillations.[13]

Figure 2a shows the measured frequencies as a function of in-plane applied field. As shown in Fig. 1b, the frequency excited depends on $I$ at a given $H$. The data in Fig. 2a correspond to the highest-frequency (lowest-current) excitation observed at a given field. Below $\mu_0 H = 50$ mT no excitations are seen. Around $\mu_0 H = 0.6$ T the excitation amplitude begins to drop and by $\mu_0 H > 1$ T is below our noise floor. The data are fit using the Kittel equation for in-plane magnon generation, excluding dipole effects, appropriate for the thin-film limit:[14]

$$f(H) = \frac{g\mu_B\mu_0}{h}\left(\left(H + \frac{Dk^2}{g\mu_B\mu_0} + H_k + M_{eff}\right)\left(H + \frac{Dk^2}{g\mu_B\mu_0} + H_k\right)\right)^{1/2} \quad (1)$$

where $D$ is the exchange stiffness, $g$ is the Landé factor, $k$ is the magnon wavenumber, $M_{eff}$ is the effective magnetization, $H_k$ is the anisotropy field, $\mu_0$ is the permeability of free space, $h$ is Planck's constant, and $\mu_B$ is the Bohr magneton. In fitting the data, $k$ and $g$ are treated as free parameters while fixed values of $\mu_0 M_{eff} = 0.8$ T and $\mu_0 H_k = 0.4$ mT are used, as determined from magnetometry measurements of analogous films. The fit yields $g = 1.78 \pm 0.01$ and an magnon wavelength of $\lambda = 390 \pm 80$ nm.  We note Eq. 1 is strictly valid only in the limit of small amplitude spin-waves, a limit not necessarily met in present measurements.  It was initially predicted that the lowest-order excited mode would have a wavelength of twice the contact diameter.[1]  However, the excitation wavelengths determined from fits to these and other data are much larger than the nominal or calculated contact sizes, which range from 25 nm to 40 nm from a Sharvin-Maxwell calculation.[15]  We infer that the excitations are ones with negligible wavevector, *i.e.* the uniform FMR mode, although this does not exclude the presence of excitations outside our measurement bandwidth.   Device-to-device variation of the measured frequency at a given $H$ is less than 10 %, while the calculated contact size varies by 60 %, consistent with the excitation of a long wavelength mode. From both the above fit and from the linear portion of the data for $\mu_0 H > 0.4$ T we determine $g = 1.78 \pm 0.01$, smaller than the value of $g = 2.0$ determined on analogous films by other methods.[12,13]

In Fig. 2b, spectra taken over a wider range of frequencies show a peak at twice the frequency of the one discussed above.  The frequencies of these peaks, along with their variations as functions of both $I$ and $H$, differ by a factor of $2.00 \pm 0.01$, with both signals being observed in fields much larger than any anisotropies in the films.  We have not observed higher-harmonic signals.  Ratios of the amplitudes of the $f$ and $2f$ signals depend on both $I$ and $H$, and show a non-monotonic dependence on $I$ and a slight increase with $H$ (inset).  For precession symmetric about the direction of the fixed layer, the frequency detected from a GMR-derived voltage should be twice the physical oscillation frequency of $M$.  A misalignment between the layers would result in the detection of a signal at $f$ as well as $2f$. Using our calculated precession angles, we estimate that a layer misalignment of a few degrees would give the $f$ to $2f$ amplitude ratios observed. Such a canting of the magnetizations may result from the applied current through either the spin-torque effect or current generated Oersted fields. Alternatively the $2f$ and $f$ signals could result from the GMR detection of the FMR mode and a period-doubled precessional oscillation, respectively. This period doubling has been observed in FMR measurements at high power (*i.e.* the Suhl instability).[16,17] In either case, the limiting slope of the data in Fig. 2a is 26 GHz/T, in good agreement with the value expected from Eq. 1 for a first harmonic signal, indicating that the lower-frequency peak corresponds to the physical precessional frequency of *M*.

These devices also emit power at lower frequencies. Figure 3 shows the $I = 5$ mA and 11 mA spectra of the device output along with the corresponding d$V$/d$I$ curve. At low currents, no signal is found. As $I$ is increased to 8 mA, a shoulder in the d$V$/d$I$ curve appears (as indicated by the two lines in the inset) and a signal is observed, the strength of which increases with current. In contrast to the high-frequency excitations, the onset of the low-frequency signals always occur at either a large peak or shoulder in the d$V$/d$I$ curve. In this device, by $I = 8$ mA the high-frequency dynamics discussed above have already turned off. However, we have measured other devices where both the high-frequency single-mode oscillations and the low-frequency signal have been simultaneously observed over a range of currents. From real-time measurements of the voltage fluctuations in these devices and nanopillars, we find that this low-frequency signal results from two-state switching in the device, as has also been reported in Ref. [5]. Shown by the $I = 10$ mA spectrum, the spectral shape follows a Lorentzian function, as expected for stochastic switching between two well defined energy states.[18] At higher currents the two-state switching typically ceases, although this is not always the case before the highest current supported by a contact (≈14 mA) is reached.

The dynamics change dramatically with applied field direction. In Fig. 4a is a two-dimensional plot showing frequency spectra as a function of $I$ for the device discussed above, but with an out-of-plane field of 0.9 T. Along the x-axis $I$ varies from 4 mA to 12 mA and back to 4 mA. A vertical slice through the plot yields a frequency spectrum at a fixed $I$. The 0.9 T field aligns the $Ni_{80}Fe_{20}$ layer with $H$ while canting the $Co_{90}Fe_{10}$ layer about 30 degrees out of the film plane. For all fields $\mu_o H > 0.6$ T, a "blue shift" in $f$ with increasing $I$ is observed. More complicated behavior is also found, e.g., jumps in the frequency occur at $I = 6$ mA and 7.5 mA. These jumps are not hysteretic, as seen from the symmetric response with increasing and decreasing $I$, and occur in all devices for out-of-plane fields. According to our modeling of this geometry, the $Ni_{80}Fe_{20}$ magnetization precesses in a nearly circular orbit about $H$, with frequency increasing with $I$, the general trend seen in our measurements. However, abrupt changes of $f$ with increasing $I$ are not found in our modeling.

These excitations persist to higher values of $H$ than for in-plane fields. As shown in Fig. 4b, dynamics persist to $\mu_o H = 1.3$ T and for $f = 38$ GHz, and can be well fit with a Lorentzian function (inset). Even at these frequencies, the linewidths are ≈ 60 MHz, and have $Q > 650$. Due to bandwidth limitations we were not able to follow the oscillations to higher frequencies. At least for point contacts, the two-state switching behavior always found with in-plane applied fields is completely suppressed in this geometry. As seen in Fig. 4b, the highest frequencies at a given $H$ vary linearly with a slope of 32

GHz/T and give $g = 2.1 \pm 0.01$, differing from the value determined from the in-plane field measurements above and the typical value obtained from other measurements. It may be that for the data shown in Fig. 4c, $H$ is not yet large enough for $f$ to be a truly linear function of $H$, leading to an inflated value of $g$, or other causes may underlie this difference. We see no other harmonics for any fields in which the excitations show a blue shift with $I$. Finally, in contrast with FMR measurements, we note that the excited frequencies discussed here increase continuously in fields ranging from $H < M_{NiFe}$ to $H > M_{NiFe}$, and persist even for $H = M_{NiFe}$, when the FMR resonance frequency is nominally zero.

In summary, we have directly measured coherent high-frequency dynamic excitations in $Co_{90}Fe_{10}/Ni_{80}Fe_{20}$ spin-valves excited by a spin-polarized DC current. The excitation frequency can be tuned over a wide range of values through both $H$ and $I$. In point contacts, this new method of inducing dynamic excitations leads to smaller values of the damping parameter than have been reported from other techniques. Single-domain simulations of current-induced dynamics qualitatively agree with many of trends measured in our devices and suggest that the current can excite a number of different precessional trajectories of the magnetization.

We thank D. C. Ralph and R. A. Buhrman for sharing unpublished results with us. We also thank P. Kabos, A. B. Kos, M. D. Stiles, and F. B. Mancoff for helpful discussions, and J. A. Katine for the nanopillar devices. This work was supported by the DARPA SPinS and NIST Nano-magnetodynamics programs.

References.

Figure Captions

FIG. 1 (a) d$V$/d$I$ vs. $I$ with $\mu_oH$ = 0.1 T. (b) High frequency spectra taken at several different values of current through the device, corresponding to the symbols in (a). Variation of $f$ with $I$ (inset).

FIG. 2 (a) Measured frequency at onset as a function of $\mu_oH$ along with fit. Error bars (FWHM) are smaller than the data points. (b) High-frequency spectra for currents from 5 mA to 9 mA taken in 0.5 mA steps with $\mu_oH$ = 0.06 T showing the fundamental response and another at twice that frequency. (inset) Amplitude ratios of the peaks as a function of $I$ for two different fields.

FIG. 3 Low-frequency spectra for two different currents along with a fit to the data at $I$ = 11 mA to a Lorentzian function. The fitted center frequency is $f_o$ = 0 ± 50 MHz. (inset) The corresponding d$V$/d$I$ curve.

FIG 4. (a) Plot of $f$ as a function of $I$ with the amplitude shown in a linear color scale from 0 (black) to 0.9 nV/Hz$^{1/2}$ (yellow). Discretization in the plot results from measuring the frequency spectra in 500 µA intervals. (b) $f$ vs. $H$ for $H$ applied out-of-plane. Data correspond to the highest $f$ at a given $H$. The error bars (FWHM) are smaller than the data points. (inset) Spectral peak at $\mu_oH$ = 1.3 T and $I$ = 11 mA along with a fit.

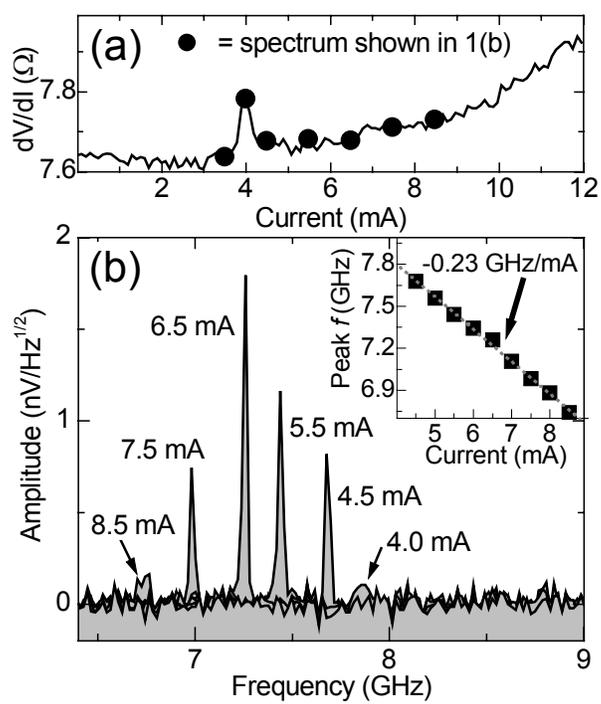

Figure 1

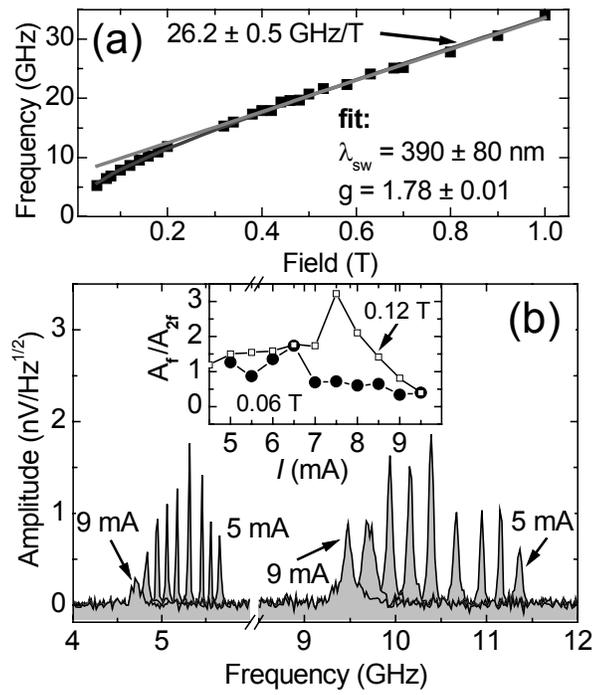

Figure 2

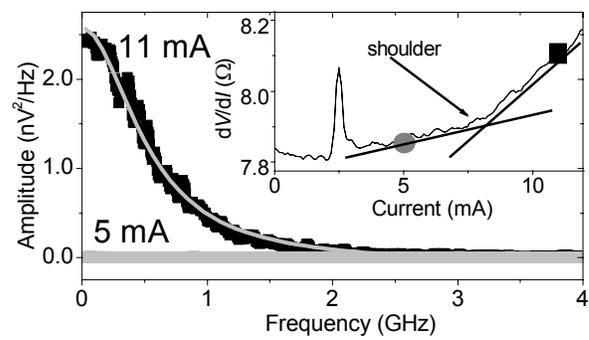

Figure 3

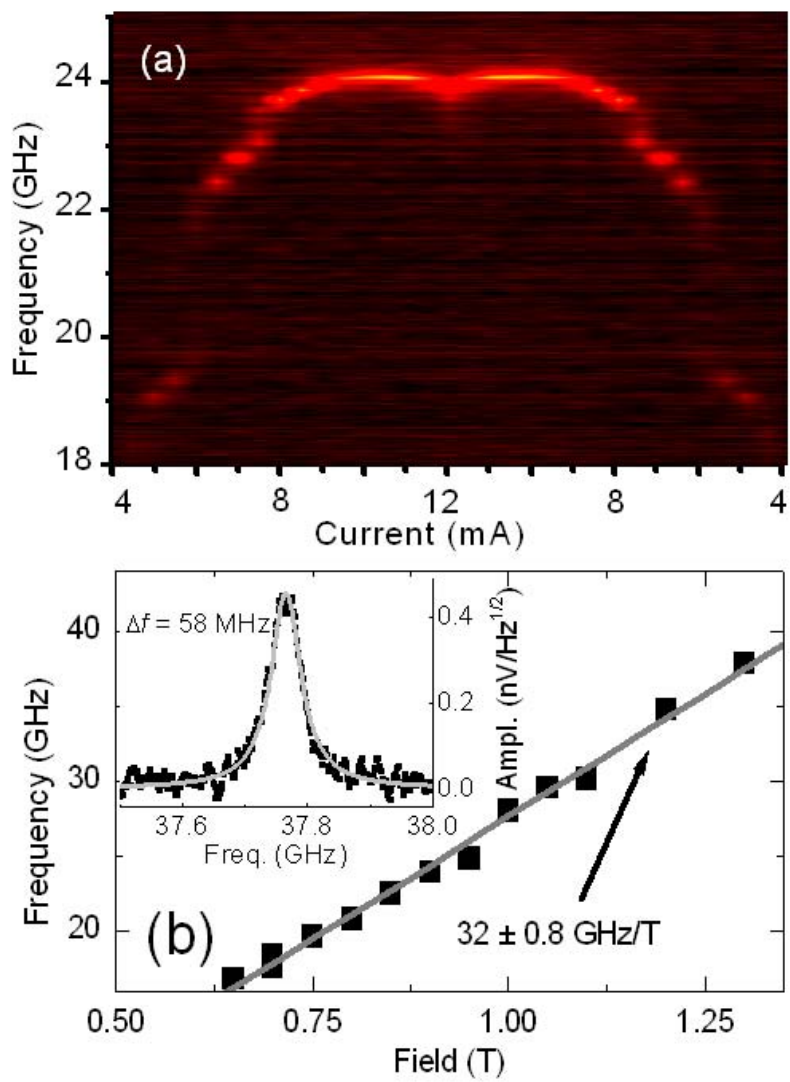

Figure 4
Rippard et al

17